\documentclass[aps,prc,superscriptaddress,reprint]{revtex4-2}
\usepackage{graphicx}
\usepackage{multirow}
\usepackage{color}
\usepackage{amssymb}
\usepackage{bbm}
\usepackage{amsmath}
\usepackage{lineno}
\usepackage{enumitem}
\usepackage{scalerel,stackengine}
\usepackage{xcolor}
\usepackage{comment}
\usepackage[normalem]{ulem}
\usepackage[colorlinks=true, breaklinks=true, linkcolor=blue, citecolor=purple, urlcolor=teal]{hyperref}

\renewcommand{\Im}{\mathop{\mathrm{Im}}}
\renewcommand{\Re}{\mathop{\mathrm{Re}}}

\def\nuc#1#2{\relax\ifmmode{}^{#1}{\protect\text{#2}}\else${}^{#1}$#2\fi}
\newcommand{\etal}{\textit{et al.~}}

\newcommand{\be}{\begin{eqnarray}}
\newcommand{\ee}{\end{eqnarray}}

\newcommand{\bwt}{\begin{widetext}}
\newcommand{\ewt}{\end{widetext}}

\stackMath
\newcommand\reallywidehat[1]{%
\savestack{\tmpbox}{\stretchto{%
  \scaleto{%
    \scalerel*[\widthof{\ensuremath{#1}}]{\kern-.6pt\bigwedge\kern-.6pt}%
    {\rule[-\textheight/2]{1ex}{\textheight}}%
  }{\textheight}%
}{0.5ex}}%
\stackon[1pt]{#1}{\tmpbox}%
}

\begin{document}

\title{Coherent Absorption Dynamics: The Dual Role of Off-Diagonal Couplings in Weakly Bound Nuclei}
\author{Hao Liu}
\author{Jin Lei}
\email[Corresponding author: ]{jinl@tongji.edu.cn}
\author{Zhongzhou Ren}%
\affiliation{%
 School of Physics Science and Engineering, Tongji University, Shanghai 200092, China.}%

\begin{abstract}
Disentangling reaction mechanisms in weakly bound nuclei remains a long-standing challenge, complicated by the common practice of treating absorption as an incoherent sum of channel contributions. Within the Continuum-Discretized Coupled-Channels (CDCC) framework, we apply the generalized optical theorem [Nucl.\ Phys.\ A \textbf{842}, 48 (2010)] and show that the total absorption cross section, $\sigma_{\mathrm{A}}\propto -\langle\Psi|W|\Psi\rangle$, decomposes as $\sigma_{\mathrm{A}}=\sigma_\mathrm{D}+\sigma_\mathrm{B}+\sigma_\mathrm{int}$, where $\sigma_\mathrm{int}$ is a coherent interference term between channel components. For the systems and complex fragment--target optical potentials considered, $\sigma_\mathrm{int}$ is negative and comparable in magnitude to the direct absorption terms. The off-diagonal imaginary couplings play a dual role: they redistribute flux among channels and generate $\sigma_\mathrm{int}$, which is required for flux-balance consistency. In calculations for $d+{}^{93}\mathrm{Nb}$ and ${}^{6}\mathrm{Li}+{}^{59}\mathrm{Co}/{}^{208}\mathrm{Pb}$, retaining the full non-diagonal coupling matrix nearly doubles the breakup-channel absorption for the heavy target, while reducing the total absorption through $\sigma_\mathrm{int}$. We demonstrate that neglecting the off-diagonal imaginary couplings $W_{ij}$ ($i\neq j$) is not merely an approximation but leads to a \emph{systematically biased physical picture}: the total absorption is overestimated while the breakup absorption component is severely underestimated. Experimental analyses that employ incoherent-sum models to extract direct and breakup cross sections from data will inherit this bias. The full coupling matrix is therefore essential for mechanism-resolved cross-section extraction, and we advocate that experimentalists adopt full-coupling CDCC calculations as the standard for consistent interpretation of absorption data in weakly bound systems.
\end{abstract}


\date{\today}

\maketitle
\section{Introduction}

Reactions involving weakly bound nuclei are of fundamental importance in nuclear astrophysics and fusion energy research, where the complex interplay between breakup and fusion processes significantly influences reaction yields~\cite{Diaz-Torres-bu-icf07,ICF18,li7Canto,jinprl19,Takada81,rangel2025nucleusnucleuspotentialsscatteringtightly,Takigawa93,Diaz-Ion02,CANTO20151,Satchler85,Udagawa85}. A central challenge in modeling these systems is the accurate disentanglement of complete fusion from other reaction channels, and the separation of absorption into direct and breakup components. Addressing this requires a theoretical framework that goes beyond evaluating total reaction probabilities to rigorously separating the absorption cross section into its constituent physical components. By adopting such a component analysis perspective, it becomes possible to explicitly examine how quantum flux is partitioned among distinct pathways and to elucidate the microscopic dynamics governing the transition from the elastic channel and continuum states to the fusion compound.

For these weakly bound systems, the Continuum-Discretized Coupled-Channels (CDCC)~\cite{AUSTERN1987125,PhysRevC.108.034612} approach provides a powerful theoretical tool to investigate the fusion-with-breakup process~\cite{ICF18,CANTO20151,li7Canto,Diaz-Ion02,Diaz-Torres-bu-icf07,Hagino_fusion00,pierre_fusion,Keely01,Diaz-ion-beck03,RANGEL2020135337,li6fusion}. By solving the coupled equations, the CDCC method describes the dynamics of the projectile within a complex nuclear potential. However, a comprehensive understanding of the reaction mechanism requires more than calculating the total cross section; it demands a clear explanation of how flux is partitioned and dissipated among different channels. Absorption cross sections are generated by the imaginary part of the optical potential, and within a coupled-channel framework, this absorption arises not only from diagonal terms but also from the coherent effects of channel-to-channel couplings.

While full-coupling calculations are computationally feasible with modern codes, simplified approaches remain prevalent in recent literature. One common approximation retains off-diagonal couplings within the continuum subspace but neglects couplings between the elastic channel and the continuum ($W_{1j}=0$ for $j>1$)~\cite{li7Canto,RANGEL2020135337,li6fusion,rangel2025nucleusnucleuspotentialsscatteringtightly}; another neglects all off-diagonal imaginary couplings entirely, retaining only diagonal terms~\cite{Diaz-Ion02,Diaz-ion-beck03}. Both approximations effectively assume an incoherent reaction picture in which absorption probabilities add. Such an assumption obscures the cross terms that arise naturally from expanding $\langle\Psi|W|\Psi\rangle$ over channel components. When adopted in the interpretation of experimental data, where model-assisted separations of direct and breakup absorption are common, this approximation can introduce a systematic bias, most notably a systematic underestimation of breakup absorption.

In this work, we apply the generalized optical theorem~\cite{COTANCH201048} within the CDCC framework to address these dynamics. The importance of off-diagonal imaginary couplings was first noted by Udagawa \etal~\cite{Udagawa85} and Satchler~\cite{Satchler85} in the context of heavy-ion coupled-channels calculations, where they identified that the absorption cross section contains an interference component from transitions between channels. Here, we apply the generalized optical theorem to obtain the complete absorption decomposition within the CDCC framework for weakly bound projectiles. This enables the theoretical separation of the total absorption cross section into three physically distinct components: direct absorption ($\sigma_\mathrm{D}$) from the elastic channel, breakup absorption ($\sigma_\mathrm{B}$) from the continuum states, and an interference term ($\sigma_\mathrm{int}$) arising from off-diagonal imaginary couplings between the elastic channel and the continuum states. Because the absorption cross section is determined by the asymptotic S-matrix through the optical theorem, this decomposition does not depend on the detailed behavior of the wave function in the nuclear interior. This theoretical separation allows us to go beyond simple incoherent summation and explicitly examine the microscopic composition of the absorption process based on the wave functions and coupling potentials.

We emphasize that ``coherence'' in this work refers specifically to the quantum-mechanical superposition of channel amplitudes during the absorption process: when the total wave function $\Psi = \sum_i \psi_i \phi_i$ is inserted into the absorption functional $\langle\Psi|W|\Psi\rangle$, cross terms $\langle\psi_i|W_{ij}|\psi_j\rangle$ appear alongside diagonal terms. The interference term $\sigma_{\rm int}$ arising from these cross terms serves as a direct quantitative measure of this coherence. This is distinct from interference between experimentally distinguishable final states, which add incoherently in any measured cross section. Crucially, a non-vanishing $\sigma_{\rm int}$ constitutes proof that the absorption mechanism involves coherent quantum interference between channels, a physical feature that incoherent summation models fundamentally fail to capture, regardless of the specific numerical value of $\sigma_{\rm int}$.

It is important to recognize that while $\sigma_\mathrm{int}$ is not directly observable as an asymptotic reaction product, its presence has concrete observable consequences. First, as we demonstrate below, neglecting the off-diagonal couplings that generate $\sigma_\mathrm{int}$ produces measurably different elastic scattering angular distributions, an observable quantity that experimentalists can directly test. Second, while the individual values of $\sigma_\mathrm{B}$ and $\sigma_\mathrm{int}$ depend on the choice of continuum discretization (bin width, shape, etc.), their sum $\sigma_\mathrm{B} + \sigma_\mathrm{int}$ is basis-independent and represents the well-defined total continuum contribution to absorption. Thus, experimentalists should care about $\sigma_\mathrm{int}$ not for its specific numerical value, but because (i) its existence invalidates incoherent-sum models as a matter of principle, and (ii) theoretical frameworks that neglect it will yield systematically biased extractions of $\sigma_\mathrm{D}$ and $\sigma_\mathrm{B}$ from data. Our numerical results demonstrate that, in the weakly bound systems studied here, $\sigma_\mathrm{int}$ is not a negligible perturbative correction but comparable in magnitude to $\sigma_\mathrm{B}$. Therefore, retaining the full coupling matrix is necessary for the correct physical interpretation of experimental data.

Adopting this perspective of component analysis, we perform a quantitative investigation of reactions induced by deuterons and lithium. Our study focuses on elucidating the dual role of the off-diagonal imaginary couplings in reaction dynamics. We demonstrate that these terms act as a "dynamical bridge," enhancing flux transfer between the ground state and the continuum, while simultaneously contributing a significant interference term that regulates the overall reaction probability. By analyzing these separated components, we show how the interference mechanism coherently modulates the total absorption and why retaining the full coupling matrix is essential for a precise evaluation of the individual contributions from direct and breakup absorption.

The paper is organized as follows. In Sec.~II, we present the theoretical formalism based on the generalized optical theorem and apply it to obtain the absorption decomposition. Section~III presents numerical results for $d+{}^{93}\mathrm{Nb}$ and ${}^{6}\mathrm{Li}+{}^{59}\mathrm{Co}/{}^{208}\mathrm{Pb}$ systems. Finally, Sec.~IV summarizes our findings and discusses their implications.

\section{Theory}
\subsection{Generalized Optical Theorem}
\label{sec:derivation}

In this subsection, we detail the generalized, coupled-channels optical theorem~\cite{COTANCH201048} and present its proof. We start from the general coupled-channels equations. The system is described by the Hamiltonian $\mathbb{H} = \hat{T}_R + \sum_{i,j} U_{ij}$, leading to the set of coupled equations for the channel wave functions $\psi_i$:
\begin{equation}
    (E_i - \hat{T}_R )\psi_i - \sum_{j=1}^N U_{ij} \psi_j = 0,
    \label{eq:cc_master}
\end{equation}
where $\hat{T}_R$ is the kinetic energy operator and $U_{ij}$ are the coupling potentials.

To simplify the structure and analyze the flux in each channel individually, we introduce the coupled-channels effective optical potential, $\bar{U}_i^{cc}$, defined by the substitution relation:
\begin{equation}
    \bar{U}_i^{cc} \psi_i \equiv \sum_{j=1}^N U_{ij} \psi_j.
    \label{eq:effective_pot_def}
\end{equation}
Substituting Eq.~\eqref{eq:effective_pot_def} into Eq.~\eqref{eq:cc_master}, the coupled system transforms into a set of formally uncoupled, single-channel equations:
\begin{equation}
    (E_i - \hat{T}_R - \bar{U}_i^{cc}) \psi_i = 0.
    \label{eq:single_channel_form}
\end{equation}
This formulation allows us to apply scattering theory techniques to each channel $i$, distinguishing between the elastic channel (with incident flux) and inelastic channels (without incident flux).

\subsubsection{Elastic Channel ($i=1$)}

For the elastic channel ($i=1$), the wave function $\psi_1$ satisfies the Lippmann-Schwinger integral equation incorporating the incident plane wave $\phi_1$:
\begin{equation}
    \psi_1 = \phi_1 + G_0^{(1)} \bar{U}_1^{cc} \psi_1,
    \label{eq:LS_elastic}
\end{equation}
where $G_0^{(1)} = (E_1 - \hat{T}_R + i\epsilon)^{-1}$ is the free Green's function. The elastic scattering amplitude is given by
\begin{equation}
    f(\theta)=-\frac{\mu}{2\pi\hbar^2}\langle\phi_1 (\mathbf{k}')|\bar{U}^{cc}_1|\psi_1 (\mathbf{k})\rangle,
\end{equation}
where $\mathbf{k}'\cdot\mathbf{k}=k_1^2\cos\theta$. The optical theorem relates the total cross section, $\sigma_\mathrm{T}$, to the forward scattering amplitude
\begin{equation}
\begin{aligned}
    \sigma_\mathrm{T} =&\frac{4\pi}{k_1}\Im f(0)\\
    =&-\frac{2\mu}{\hbar^2 k_1} \Im \langle \phi_1 | \bar{U}_1^{cc} | \psi_1 \rangle.
\end{aligned}
    \label{eq:sigma_total_def}
\end{equation}
To isolate the reaction cross section, we substitute the bra-vector form of Eq.~\eqref{eq:LS_elastic}, $\langle \phi_1 | = \langle \psi_1 | - \langle \psi_1 | \bar{U}_1^{cc \dagger} G_0^{(1)\dagger}$, into Eq.~\eqref{eq:sigma_total_def}. This decomposes the total cross section into shape elastic ($\sigma_\mathrm{SE}$) and reaction ($\sigma_\mathrm{R}$) components:
\begin{equation}
    \sigma_\mathrm{T} = \underbrace{\frac{2\mu}{\hbar^2 k_1} \Im \langle \psi_1 | \bar{U}_1^{cc \dagger} G_0^{(1)\dagger} \bar{U}_1^{cc} | \psi_1 \rangle}_{\sigma_\mathrm{SE}} 
    \underbrace{- \frac{2\mu}{\hbar^2 k_1} \Im \langle \psi_1 | \bar{U}_1^{cc} | \psi_1 \rangle}_{\sigma_\mathrm{R}}.
\end{equation}
The reaction cross section $\sigma_\mathrm{R}$ is thus generated solely by the expectation value of the imaginary part of the effective potential:
\begin{equation}
    \sigma_\mathrm{R} = -\frac{2}{\hbar v_1} \Im \langle \psi_1 | \bar{U}_1^{cc} | \psi_1 \rangle.
    \label{eq:sigma_R_final}
\end{equation}
\subsubsection{Inelastic Channels (\texorpdfstring{$i \neq 1$}{i ≠ 1})}
For inelastic or breakup channels ($i \neq 1$), there is no incident plane wave ($\phi_i = 0$). The Lippmann-Schwinger equation becomes:
\begin{equation}
    \psi_i = G_0^{(i)} \sum_{j=1}^N U_{ij} \psi_j = G_0^{(i)} \bar{U}_i^{cc} \psi_i.
    \label{eq:LS_inelastic}
\end{equation}
The standard definition of the cross section for channel $i$ is the integral of the scattering amplitude squared:
\begin{equation}
    \sigma_i = \frac{v_i}{v_1} \int d \Omega_i \left| f_i(\theta_i) \right|^2 
    = \frac{v_i}{v_1} \int d \Omega_i \left| \frac{\mu}{2 \pi \hbar^2}\langle\phi_i | \bar{U}_i^{cc} | \psi_i \rangle \right|^2.
    \label{eq:sigma_i_std}
\end{equation}
Alternatively, using the generalized optical theorem~\cite{COTANCH201048}, the cross section can be more simply calculated from:
\begin{equation}
    \sigma_i = \frac{2}{\hbar v_1} \Im \langle \psi_i | \bar{U}_i^{cc} | \psi_i \rangle.
    \label{eq:sigma_i_prop}
\end{equation}
Note the sign difference between Eq.~\eqref{eq:sigma_i_prop} and Eq.~\eqref{eq:sigma_R_final}: the elastic channel carries a minus sign, whereas the inelastic channels do not. This difference reflects the distinct physical content of $\Im\langle\psi_i|\bar{U}_i^{cc}|\psi_i\rangle$ in each case. For the elastic channel ($i=1$), the effective potential is dominated by the diagonal absorption term $U_{11}$, whose imaginary part $W_{11}<0$ removes flux from the incident wave; hence $\Im\langle\psi_1|\bar{U}_1^{cc}|\psi_1\rangle < 0$, and a minus sign is required to yield a positive cross section. For inelastic channels ($i\neq 1$), however, no incident wave exists; the wave function $\psi_i$ is generated entirely by coupling from the elastic channel. The dominant contribution to $\bar{U}_i^{cc}\psi_i = \sum_j U_{ij}\psi_j$ comes from the source term $U_{i1}\psi_1$ rather than the diagonal absorption $U_{ii}\psi_i$, because $|\psi_1| \gg |\psi_i|$. As shown in the proof below, this source-dominated structure leads to $\Im\langle\psi_i|\bar{U}_i^{cc}|\psi_i\rangle > 0$, so no additional sign is needed. \textit{This observation already hints at the central role of ground-state--continuum couplings: neglecting $U_{i1}$ would starve the breakup channels of their primary flux source.}

For completeness, we present the proof of Eq.~\eqref{eq:sigma_i_prop} as published in Ref.~\cite{COTANCH201048}. Starting from the right-hand side of Eq.~\eqref{eq:sigma_i_prop}, we substitute the ket-vector $\psi_i$ using the LS equation (Eq.~\eqref{eq:LS_inelastic}):
\begin{equation}
    \Im \langle \psi_i | \bar{U}_i^{cc} | \psi_i \rangle 
    = \Im \langle G_0^{(i)} \bar{U}_i^{cc}\psi_i | \bar{U}_i^{cc} |  \psi_i \rangle 
    = \Im \langle \Phi_i | G_0^{(i)\dagger} | \Phi_i \rangle,
\end{equation}
where we defined the source term $|\Phi_i\rangle = \bar{U}_i^{cc} |\psi_i\rangle$. Using the operator identity for the Green's function, $\Im G_0^{(i)\dagger} = \pi \delta(E_i - \hat{H}_0)$, and inserting a complete set of plane wave states $\{|\phi_i(\mathbf{k}_i')\rangle\}$, we obtain:
\begin{align}
    \Im \langle \Phi_i | G_0^{(i)\dagger} | \Phi_i \rangle
    &= \pi \int \frac{d\mathbf{k}_i'}{(2\pi)^3} \langle \Phi_i | \phi_i(\mathbf{k}_i') \rangle \delta(E_i - E_i') \langle \phi_i(\mathbf{k}_i') | \Phi_i \rangle \nonumber \\
    &= \pi \frac{\mu k_i}{(2\pi)^3 \hbar^2} \int d\Omega_i \left| \langle \phi_i | \bar{U}_i^{cc} | \psi_i \rangle \right|^2.
\end{align}
Substituting this result back into Eq.~\eqref{eq:sigma_i_prop}, the prefactors cancel perfectly to yield:
\begin{equation}
    \sigma_i = \frac{2}{\hbar v_1} \left( \pi \frac{\mu k_i}{(2\pi)^3 \hbar^2} \int d\Omega_i \left| \langle \phi_i | \bar{U}_i^{cc} | \psi_i \rangle \right|^2 \right)
    = \frac{v_i}{v_1} \int d\Omega_i |f_i(\theta_i)|^2.
\end{equation}
This confirms that the compact form of Eq.~\eqref{eq:sigma_i_prop} is mathematically equivalent to the standard cross section definition (Eq.~\eqref{eq:sigma_i_std}). Beyond its theoretical elegance, this form is computationally advantageous as it enables the calculation of partial cross sections directly from wave function overlaps, bypassing the need for complex asymptotic integrations.

Physically, this result shows the flux dynamics within inelastic channels ($i \neq 1$), where no incident wave exists. In these channels, $\sigma_i$ quantifies the net flux emerging from the reaction. It represents the accumulated flux populated by coupling source terms from other channels, accounting for any subsequent reduction caused by the imaginary part of the coupling potential (absorption).

\subsection{Cross Section Analysis within CDCC}
\label{sec:cross_section_analysis}

We now apply the general formalism presented in Sec.~\ref{sec:derivation} to the specific physical problem of projectile breakup. The three-body Hamiltonian describing the collision is
\begin{equation}
\mathbb{H}(\mathbf{R},\mathbf{r}) = h(\mathbf{r}) + \hat{K}_R + \mathbb{U}^{(1)}(r_1) + \mathbb{U}^{(2)}(r_2),
\end{equation}
where $\mathbf{R}$ describes the projectile-target relative motion, and $\mathbf{r}$ the projectile internal structure. The fragment-target interactions are complex optical potentials, $\mathbb{U}^{(k)} = V^{(k)} + i W^{(k)}$.

In the Continuum-Discretized Coupled-Channels (CDCC) framework~\cite{AUSTERN1987125,Thompson_Nunes_2009,PhysRevC.108.034612}, the total wavefunction is expanded in a truncated basis of projectile states $\{\phi_i(\mathbf{r})\}$. This leads to the coupled equations:
\begin{equation}
\left(E - \epsilon_i - \hat{T}_R\right) \psi_i(\mathbf{R}) 
= \sum_{j=1}^N U_{ij}(\mathbf{R})\,\psi_j(\mathbf{R}),
\label{eq:CDCC}
\end{equation}
where the coupling potentials are explicitly defined as:
\begin{equation}
U_{ij}(\mathbf{R}) = \langle \phi_i | \mathbb{U}^{(1)} + \mathbb{U}^{(2)} | \phi_j \rangle 
= V_{ij}(\mathbf{R}) + i W_{ij}(\mathbf{R}).
\label{eq:coulped_pot}
\end{equation}
Here, channel $i=1$ denotes the elastic channel.

Based on the reaction cross section obtained for the elastic channel (Eq.~\eqref{eq:sigma_R_final}), the total flux loss from the elastic channel, $\sigma_\mathrm{R}$, is determined by the imaginary part of the couplings connecting to all channels:
\begin{align}
\sigma_\mathrm{R} &= -\frac{2}{\hbar v_1} \Im \langle \psi_1 | \bar{U}_1^{cc} | \psi_1 \rangle \nonumber\\
&= -\frac{2}{\hbar v_1} \Im \left[ \sum_{j=1}^N \langle \psi_1 | i W_{1j} | \psi_j \rangle 
+ \sum_{j=2}^N \langle \psi_1 | V_{1j} | \psi_j \rangle \right].
\label{eq:sigmaR}
\end{align}

The flux transferred into the breakup channels (denoted as $i \ge 2$) constitutes the elastic breakup (EBU) cross section. Using the result from the generalized optical theorem for inelastic channels (Eq.~\eqref{eq:sigma_i_prop}), $\sigma_{\rm EBU}$ is obtained by summing the partial cross sections $\sigma_i$:
\begin{align}
\sigma_{\rm EBU} &= \sum_{i=2}^N \sigma_i
= \frac{2}{\hbar v_1} \Im \sum_{i=2}^N \langle \psi_i | \sum_{j=1}^N U_{ij} \psi_j \rangle \nonumber\\
&= \frac{2}{\hbar v_1} \Im \left[ \sum_{i=2}^N \langle \psi_i | V_{i1} | \psi_1 \rangle
+ \sum_{i=2}^N \langle \psi_i | \sum_{j=1}^N i W_{ij} | \psi_j \rangle \right].
\label{eq:sigmaEBU}
\end{align}

Within the CDCC framework, elastic scattering and breakup are treated explicitly, while processes such as fusion, transfer, and target excitation are modeled implicitly through the imaginary potentials $W_{ij}$. The total absorption cross section $\sigma_\mathrm{A}$ quantifies the flux removed by these implicit channels and is defined as the difference between the total reaction cross section and the elastic breakup cross section:
\begin{equation}
\sigma_\mathrm{A} = \sigma_\mathrm{R} - \sigma_{\rm EBU}.
\end{equation}
Substituting Eqs.~\eqref{eq:sigmaR} and \eqref{eq:sigmaEBU} into this expression, the terms involving the real coupling potentials $V_{ij}$ cancel out due to the Hermiticity of the potential matrix ($\langle \psi_i | V_{ij} | \psi_j \rangle = \langle \psi_j | V_{ji} | \psi_i \rangle^*$). We are left with the compact result:
\begin{equation}
\begin{aligned}
\sigma_\mathrm{A} =& -\frac{2}{\hbar v_1}
\Im \sum_{i=1}^N \sum_{j=1}^N \langle \psi_i | i W_{ij} | \psi_j \rangle \\
=& -\frac{2}{\hbar v_1} \sum_{i=1}^N \sum_{j=1}^N \langle \psi_i | W_{ij} | \psi_j \rangle.
\end{aligned}
\label{eq:sigmaA_final}
\end{equation}
The second equality follows because the double sum is purely real: for symmetric couplings ($W_{ij}=W_{ji}$), each pair $(i,j)$ and $(j,i)$ contributes $\langle\psi_i|W_{ij}|\psi_j\rangle + \langle\psi_j|W_{ji}|\psi_i\rangle = 2\Re\langle\psi_i|W_{ij}|\psi_j\rangle$, so the total sum equals its real part.

Equation~\eqref{eq:sigmaA_final} demonstrates that $\sigma_\mathrm{A}$ is the coherent sum of expectation values of the imaginary coupling potentials. A similar decomposition was noted by Udagawa \etal~\cite{Udagawa85} and Satchler~\cite{Satchler85} using flux-balance arguments in the context of heavy-ion fusion. Our application of the generalized optical theorem (Sec.~\ref{sec:derivation}) provides a self-contained proof based on the Lippmann-Schwinger equation and the completeness of plane-wave states, and extends the result to the CDCC framework for weakly bound projectiles.

It is important to distinguish between two types of imaginary potentials that appear in the literature. In the present work, the imaginary potentials $W^{(k)}$ are standard phenomenological optical potentials constrained by fits to elastic scattering data, representing the dynamical polarization potential (DPP) arising from the Feshbach projection of all channels excluded from the CDCC model space. These potentials encompass \emph{all} flux removed from the model space, including not only fusion but also transfer reactions and other non-elastic processes. An alternative approach, adopted in some fusion calculations~\cite{Hagino_fusion00,Diaz-Ion02}, employs either the ingoing wave boundary condition (IWBC) or a short-range, strongly absorbing imaginary potential added \textit{ad hoc} to the Hamiltonian to simulate the complete absorption of flux in the inner region of the barrier. In such calculations, the off-diagonal matrix-elements of the short-range fusion potential involve wave functions in a region where the model Hamiltonian may not be reliable; consequently, the resulting transitions may lack clear physical meaning. In the present work, the optical potentials are constrained by elastic scattering data and the absorption cross section is determined by the asymptotic S-matrix through the optical theorem. The conclusions of this work therefore concern the model-space absorption and should not be directly compared with CF/ICF cross sections obtained from IWBC or short-range fusion potentials.

Consequently, the quantities $\sigma_\mathrm{D}$ (direct) and $\sigma_\mathrm{B}$ (breakup) defined below should be understood as components of the \emph{total absorption}, not exclusively fusion. \textit{Mapping condition:} $\sigma_\mathrm{D}$ and $\sigma_\mathrm{B}$ may be interpreted as proxies for complete and incomplete fusion only when the imaginary potential is specifically designed to represent absorption leading to fusion (e.g., a short-range, strongly absorbing potential localized inside the Coulomb barrier). With standard phenomenological optical potentials, as used in the present work, these quantities represent the more general ``direct-channel absorption'' and ``breakup-channel absorption.'' Despite this caveat, the central result of this work, that coherent interference generates a significant $\sigma_\mathrm{int}$ term, remains valid for any absorption mechanism described by non-diagonal imaginary couplings, including transfer.

To explicitly examine the flux partitioning and the role of quantum interference, we decompose this coherent sum into three physically distinct components:
\begin{equation}
    \sigma_\mathrm{A} = \sigma_\mathrm{D} + \sigma_\mathrm{B} + \sigma_\mathrm{int},
    \label{eq:sigmaA_complete}
\end{equation}
where $\sigma_\mathrm{D}$ and $\sigma_\mathrm{B}$ represent the diagonal-like absorption from the elastic and continuum channels, respectively.

\textit{Basis dependence.} We note that while the total absorption $\sigma_\mathrm{A}$ is basis-independent, the partition into $\sigma_\mathrm{D}$, $\sigma_\mathrm{B}$, and $\sigma_\mathrm{int}$ depends on the choice of channel basis within the continuum subspace. However, two important quantities are basis-independent: (i) $\sigma_\mathrm{D}$, because the elastic channel (ground state) is uniquely defined by the physical projectile, and (ii) the sum $\sigma_\mathrm{B} + \sigma_\mathrm{int}$, representing the total continuum contribution.\footnote{To see this, let $P_g$ project onto the ground state and $P_c = 1 - P_g$ onto the continuum subspace. Then $\sigma_\mathrm{A} \propto -\langle\Psi|W|\Psi\rangle$ and $\sigma_\mathrm{D} \propto -\langle\Psi|P_g W P_g|\Psi\rangle$. The remainder, $\sigma_\mathrm{A} - \sigma_\mathrm{D} \propto -\langle\Psi|(P_c W P_c + P_g W P_c + P_c W P_g)|\Psi\rangle$, is invariant under any unitary transformation $U$ acting within the continuum subspace ($P_c \to U P_c U^\dagger$), since both the basis states and the operator transform together, leaving the expectation value unchanged.} A unitary rotation among continuum bins redistributes flux between $\sigma_\mathrm{B}$ and $\sigma_\mathrm{int}$ individually but leaves their sum invariant. The key physical conclusion, that neglecting $W_{1j}$ modifies both observables (elastic scattering) and the total absorption, is therefore robust under basis changes.

The direct absorption component represents absorption from the elastic channel: 
\begin{equation} 
\sigma_\mathrm{D} = -\frac{2}{\hbar v_1} \langle \psi_1 | W_{11} | \psi_1 \rangle, 
\label{eq:D}
\end{equation} 
and the breakup absorption component represents absorption from the continuum states:
\begin{equation}
\sigma_\mathrm{B} = -\frac{2}{\hbar v_1} \sum_{i=2}^N \sum_{j=2}^N \langle \psi_i | W_{ij} | \psi_j \rangle.
\label{eq:B}
\end{equation} 
The third term, $\sigma_\mathrm{int}$, represents the interference contribution arising from the off-diagonal imaginary couplings connecting the ground state to the continuum:
\begin{equation}
    \sigma_\mathrm{int} = -\frac{2}{\hbar v_1} \left[ \sum_{j=2}^N \langle \psi_1 | W_{1j} | \psi_j \rangle + \sum_{i=2}^N \langle \psi_i | W_{i1} | \psi_1 \rangle \right].
    \label{eq:sigmaint_general}
\end{equation}
For symmetric couplings ($W_{ij} = W_{ji}$), which hold for fragment--target optical potentials derived from the same nucleon--target interaction, this simplifies to:
\begin{equation}
    \sigma_\mathrm{int} = -\frac{4}{\hbar v_1} \sum_{j=2}^N \Re\langle \psi_1 | W_{1j} | \psi_j \rangle.
    \label{eq:sigmaint}
\end{equation}
In both forms, $\sigma_\mathrm{int}$ is manifestly real. Its sign depends on the relative phases of the channel wave functions, which are determined by the coupled-channel dynamics.

The decomposition in Eq.~(\ref{eq:sigmaA_final}) highlights that all imaginary parts of the coupling potentials, both diagonal and off-diagonal, contribute coherently to the genuine absorption. If the fragment-target interactions are real ($W^{(1)}=W^{(2)}=0$), then $\sigma_\mathrm{A}=0$ and $\sigma_\mathrm{R} = \sigma_{\rm EBU}$, as expected for pure breakup with no flux loss.

With different models, different imaginary potentials have been used in CDCC calculations. It is useful to distinguish the two main approaches.

In the first approach, adopted by Diaz-Torres and Thompson~\cite{Diaz-Ion02} and Hagino \etal~\cite{Hagino_fusion00}, the imaginary potential is a short-range, strongly absorbing potential added to the diagonal terms to simulate fusion by absorbing all flux that reaches the inner region of the barrier. (Equivalently, the ingoing wave boundary condition (IWBC) may be applied at an inner radius to achieve the same effect~\cite{Hagino_fusion00}.) In this case, the real interaction generates the coupling potential, while the imaginary couplings are diagonal in the channel index ($W_{ij}=0$ for $i \neq j$). The interference term $\sigma_\mathrm{int}$ then vanishes, and Eq.~\eqref{eq:sigmaA_complete} reduces to an incoherent sum:
\begin{equation}
\sigma_\mathrm{A} = -\frac{2}{\hbar v_1} \sum_{i=1}^N \langle \psi_i | W_{ii} | \psi_i \rangle. \label{eq:sigmaA_diag}
\end{equation}
The corresponding direct and breakup components are given by Eqs.~\eqref{eq:D} and \eqref{eq:B}.

In the second approach, adopted in other CDCC calculations~\cite{li7Canto,li6fusion,pierre_fusion}, the fragment--target interactions include phenomenological optical potentials $\mathbb{U}^{(k)} = V^{(k)} + i W^{(k)}$ that are constrained by elastic scattering data. Because the imaginary potentials $W^{(k)}$ act on the fragment--target coordinates, they generate non-vanishing off-diagonal matrix-elements $W_{ij}$ between all channel pairs when projected onto the CDCC basis (Eq.~\eqref{eq:coulped_pot}). This is the approach we follow in the present work.

In many recent studies investigating fusion mechanisms~\cite{ICF18,li7Canto,li6fusion}, it is common to neglect these off-diagonal imaginary couplings ($W_{1j}$ and $W_{j1}$, we denote this approximation as $W_{1j}=0$) between the ground state and continuum states in Eq.~\eqref{eq:coulped_pot}. Under this approximation, the absorption cross section is simplified to:
\begin{equation}
\sigma_\mathrm{A} = \sigma_\mathrm{D} + \sigma_\mathrm{B}, 
\label{eq:fusion}
\end{equation} 
where $\sigma_\mathrm{D} = -\frac{2}{\hbar v_1} \langle \psi_1 | W_{11} | \psi_1 \rangle$  and $\sigma_\mathrm{B}$ is the absorption from continuum states. This model implies that absorption events are localized strictly within individual channels, treating the coupling terms purely as flux transporters with no 'absorption' generated during the transition itself. However, as we demonstrate in the subsequent section, this omission renders the physical description incomplete, as $\sigma_\mathrm{int}$ provides a significant coherent contribution to the total absorption.

\subsection{Dynamical polarization potential and a coupling-induced absorption measure}
\label{subsec:DPP}

The coupled-channel equations \eqref{eq:cc_master} can be rewritten in a single-channel form by introducing the channel-dependent effective (or \emph{polarization}) potential,
\begin{equation}
\bar{U}^{cc}_i\,\psi_i \equiv \sum_{j=1}^{N}U_{ij}\psi_j,
\label{eq:Ucc_def_again}
\end{equation}
so that
\begin{equation}
\left(E_i-\hat{T}_R-\bar{U}^{cc}_i\right)\psi_i=0 .
\label{eq:single_channel_again}
\end{equation}
For the elastic channel ($i=1$), it is natural to separate $\bar{U}^{cc}_1$ into the bare elastic potential and an induced term,
\begin{equation}
\bar{U}^{cc}_1 = U_{11} + \Delta U_{\rm DPP},
\label{eq:DPP_split}
\end{equation}
where $\Delta U_{\rm DPP}$ is the dynamical polarization potential (DPP) generated by couplings to the breakup (continuum) subspace. In practice, $\Delta U_{\rm DPP}$ may be constructed via Feshbach projection~\cite{liu2025exact}, or through its trivially equivalent local potential (TELP)~\cite{Thompson_Nunes_2009} representation; the discussion below is independent of the specific construction, as it relies only on the identity \eqref{eq:DPP_split}.

Using the optical-theorem form for the reaction cross section in the elastic channel, Eq.~\eqref{eq:sigma_R_final}, one obtains
\begin{equation}
\sigma_R
= -\frac{2}{\hbar v_1}\Im\langle\psi_1|\bar{U}^{cc}_1|\psi_1\rangle
= \sigma_D + \sigma_{\rm DPP},
\label{eq:sigmaR_DPP}
\end{equation}
with
\begin{equation}
\begin{aligned}
\sigma_D &\equiv -\frac{2}{\hbar v_1}\Im\langle\psi_1|U_{11}|\psi_1\rangle,
\\
\sigma_{\rm DPP} &\equiv -\frac{2}{\hbar v_1}\Im\langle\psi_1|\Delta U_{\rm DPP}|\psi_1\rangle .
\end{aligned}
\label{eq:sigmaD_sigmaDPP_def}
\end{equation}
Here \(\sigma_D\) represents absorption associated with the bare elastic optical potential, whereas \(\sigma_{\rm DPP}\) isolates the additional reaction flux induced by coupling to the continuum. Therefore, in the present work we use \(\sigma_{\rm DPP}\) as a quantitative measure of the net influence of coupled-channel (continuum) effects on reaction absorption.

The connection between \(\sigma_{\rm DPP}\) and the channel-resolved decomposition in Sec.~\ref{sec:cross_section_analysis} follows from \(\sigma_R=\sigma_{\rm EBU}+\sigma_A\), together with Eq.~\eqref{eq:sigmaA_complete}:
\begin{equation}
\sigma_{\rm DPP}
= \sigma_R-\sigma_D
= \sigma_{\rm EBU} + \sigma_A - \sigma_D
= \sigma_{\rm EBU} + \sigma_B + \sigma_{\rm int}.
\label{eq:sigmaDPP_relation}
\end{equation}
Equation~\eqref{eq:sigmaDPP_relation} clarifies the physical content of \(\sigma_{\rm DPP}\). When the fragment--target interactions are real, \(\sigma_A=0\) and \(\sigma_{\rm DPP}\) reduces to the purely elastic breakup cross section. With complex optical potentials, however, \(\sigma_{\rm DPP}\) contains not only elastic breakup but also continuum-induced absorption, encoded by \(\sigma_B\) and the coherent interference contribution \(\sigma_{\rm int}\) generated by the off-diagonal imaginary couplings. This is precisely why \(\sigma_{\rm DPP}\) provides a convenient and robust diagnostic of coupling effects in our calculations: it quantifies the additional reaction flux attributable to the inclusion of the continuum beyond the bare elastic absorption.

\section{Results}

\subsection{Role of Off-Diagonal Imaginary Couplings: d+$^{93}$Nb Test Case}

To quantitatively assess the necessity of retaining the full non-diagonal imaginary coupling matrix $W_{ij}$, specifically the terms connecting the ground state to continuum bins, we first perform a controlled numerical experiment using the deuteron breakup reaction d+$^{93}$Nb at $E_{\text{lab}} = 25.5$ MeV. This system serves as an ideal test case due to its simple two-body cluster structure ($d=p+n$) and manageable model space. We introduce a numerical flag in the CDCC and demonstrate the impact by comparing the full calculation, where all matrix elements $U_{ij} = V_{ij} + iW_{ij}$ are retained, against an approximation where the imaginary couplings between the ground state and continuum are neglected ($W_{1j} = W_{j1} = 0$ for $j \geq 2$, denoted $W_{1j}=0$ below). In this toy model, we use the same model space: the $p-n$ continuum states are only on $l=0$ partial wave, using a simple Gaussian interaction as discussed in Ref.~\cite{AUSTERN1987125}. The energy bins are extended to 12~MeV which divided into 6 bin states. The nucleon-target interactions are taken from KD02~\cite{KD02}. To simplify the calculation, the spins of the particles are ignored, and closed channels are not included in the model space.

\begin{figure}[t]
    \centering
    \includegraphics[width=1.0\linewidth]{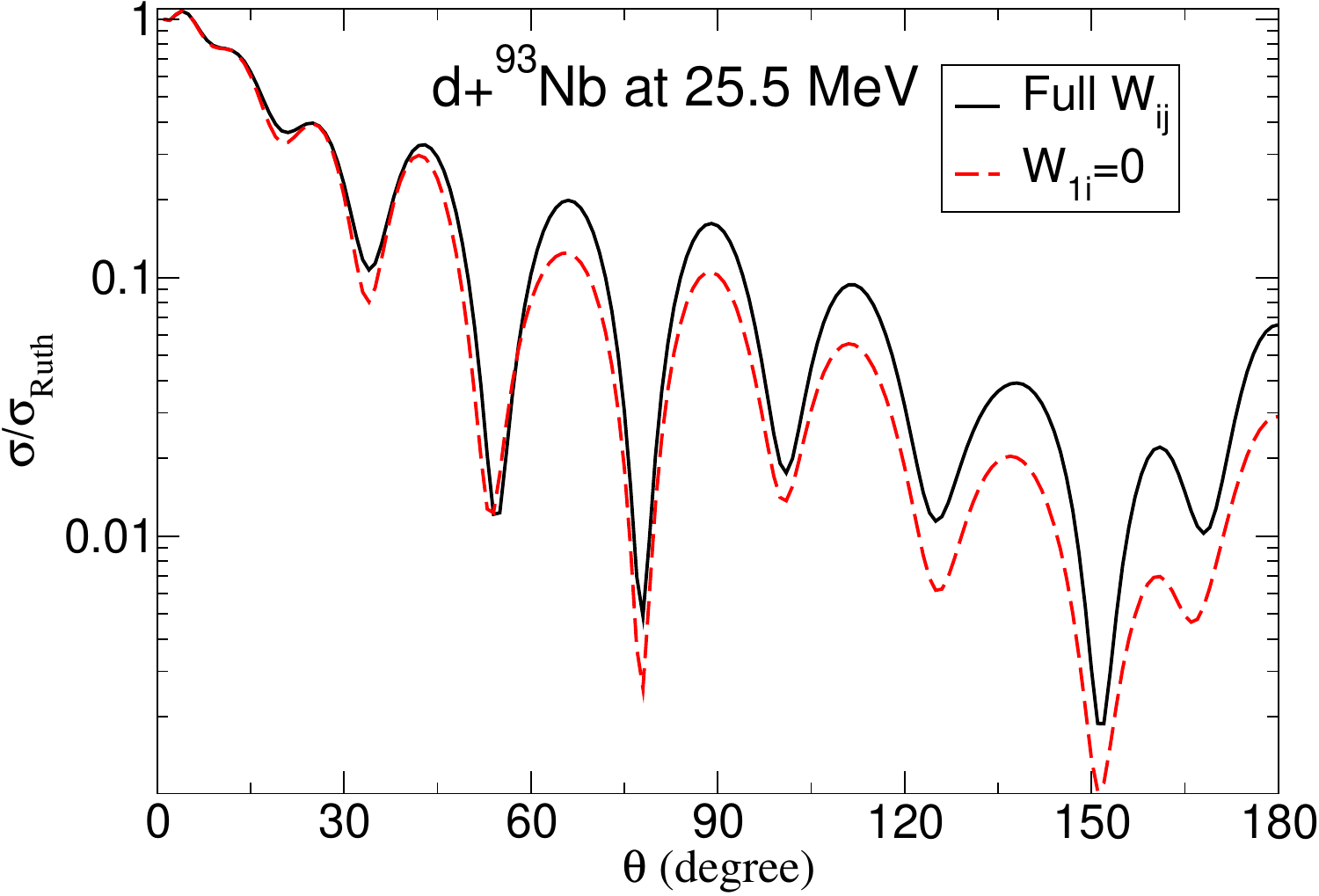}
    \caption{Elastic scattering cross sections from CDCC calculations (comparing the full calculation and the approximation with $W_{1j}=0$) for the $d + ^{93}\mathrm{Nb}$ reaction at 25.5 MeV.}
    \label{fig:elastic}
\end{figure}

Figure~\ref{fig:elastic} presents the elastic scattering cross section distributions from both the full calculation (solid black line) and the approximation with $W_{1j}=0$ (dashed red line). It is important to note that, even with the relatively simple coupling scheme employed here, significant deviations are observed between the two results, particularly in the intermediate-angle region ($30^\circ - 120^\circ$). These differences are not merely quantitative; they reflect fundamentally different scattering dynamics.

In the full calculation, the off-diagonal imaginary couplings $W_{1j}$ (coupling the elastic channel to continuum bins) allow coherent flux redistribution through quantum interference effects. When these couplings are artificially set to zero, the scattering amplitude loses this interference mechanism, resulting in a systematically different angular distribution. The most striking differences occur around $\theta \approx 60^\circ$ and $120^\circ$, where the full calculation exhibits deeper minima compared to the approximation. This leads to a crucial observation: although $\sigma_\mathrm{int}$ itself is not an asymptotic observable, neglecting it produces measurable consequences in the elastic channel. Any theoretical framework that omits the off-diagonal imaginary couplings will fail to reproduce the correct elastic angular distribution, providing experimentalists with a direct, testable criterion for the necessity of retaining the full coupling matrix.

To facilitate the discussion of flux dynamics, we analyze the calculated absorption cross sections using the coherent decomposition defined in Eq.~(\ref{eq:sigmaA_complete}). As shown in Table~\ref{tab:w1j_impact}, the interference term $\sigma_\mathrm{int}$ represents the coupling effect between the elastic and continuum channels. When the ground-state-to-continuum imaginary couplings are artificially neglected, $\sigma_\mathrm{int}$ vanishes by definition, and the total absorption reduces to the incoherent sum (Eq.~\eqref{eq:fusion}).


\begin{figure*}[htbp]
    \centering
    \includegraphics[width=1.0\linewidth]{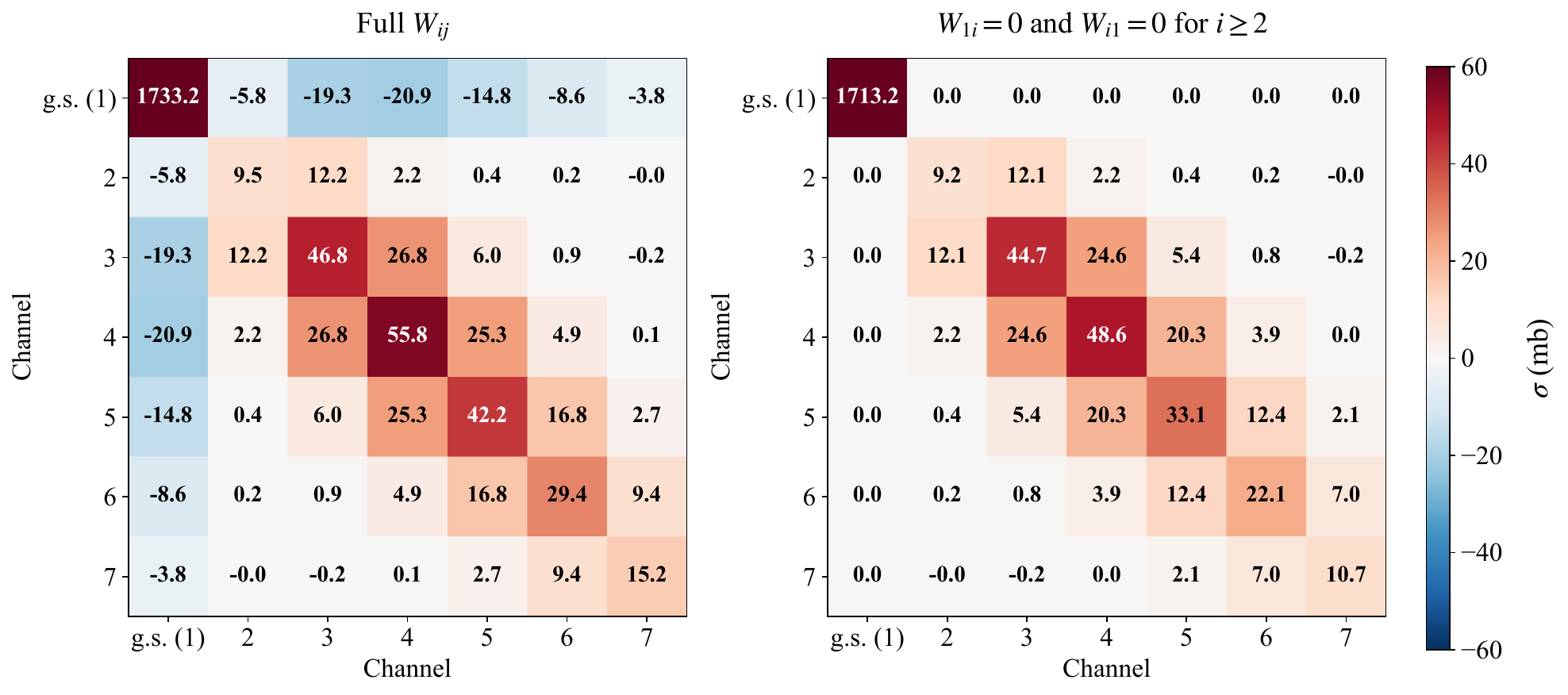}
    \caption{Comparison of absorption cross-section matrices for d+$^{93}$Nb at 25.5 MeV ($l=0$ bins). \textbf{Left}: Full coupling calculation. \textbf{Right}: Approximation with $W_{1j}=0$. The negative off-diagonal values in the full calculation (left) indicate coherent interference flux loss, which is artificially removed in the approximation (right).}
    \label{fig:w1j_comparison}
\end{figure*}

\begin{table}[htbp]
\centering
\caption{Impact of neglecting $W_{1j}$ on absorption cross sections for d+$^{93}$Nb ($l=0$ bins). The "Diff." column represents (Full $W_{ij}$ $-$ $W_{1j}=0$).}
\label{tab:w1j_impact}
\begin{tabular}{lccc}
\hline\hline
Quantity & Full $W_{ij}$ & $W_{1j}=0$ & Diff. \\
\hline
$\sigma_\mathrm{D}$ [mb] & 1733.2 & 1713.2 & $+20.0$ \\
$\sigma_\mathrm{B}$ [mb] & 414.1 &  350.8 & $+63.3$ \\
$\sigma_\mathrm{int}$ [mb] & $-146.7$ & $0.0$ & $-146.7$ \\
$\sigma_\mathrm{A}$ [mb] & 2000.7 & 2064.0 & $-63.3$ \\
\hline\hline
\end{tabular}
\end{table}

The comparison of absorption cross-section matrices, shown in Figure~\ref{fig:w1j_comparison}, reveals that the interference terms in the full calculation exhibit significant negative values. As shown in Eq.~\eqref{eq:sigmaA_final}, these negative contributions in the off-diagonal blocks are a direct signature of the quantum interference between the elastic and continuum channels. Table~\ref{tab:w1j_impact} summarizes the quantitative impact of neglecting these terms. In the full calculation, the interference term is substantial and negative ($\sigma_\mathrm{int} = -146.7$ mb), which acts to reduce the total absorption cross section via destructive interference. When we artificially set $\sigma_\mathrm{int} = 0$ in the approximation, the flux distribution is fundamentally altered. Contrary to naive expectations, the total absorption cross section in the approximation ($\sigma_\mathrm{A} = 2064.0$ mb) is actually larger than in the full calculation ($\sigma_\mathrm{A} = 2000.7$ mb). This counterintuitive result occurs because the approximation removes the large negative interference term that naturally regulates the total absorption in the full physical picture. Simultaneously, the removal of the coupling leads to a significant underestimation of the breakup absorption component, with $\sigma_\mathrm{B}$ dropping from 414.1 mb to 350.8 mb, and a slight decrease in direct absorption $\sigma_\mathrm{D}$ from 1733.2 mb to 1713.2 mb.

This comparison demonstrates that absorption involves a coherent superposition where the interference couplings play a dual role: they serve to enhance the specific absorption pathways ($\sigma_\mathrm{B}$ and $\sigma_\mathrm{D}$) by redistributing flux between channels, while simultaneously reducing the total reaction probability through destructive interference. Consequently, the approximation of neglecting off-diagonal imaginary couplings fails on two fronts: it overestimates the total absorption while underestimating the individual absorption components. This implies that prior analyses based on incoherent approximations likely underestimated breakup absorption cross sections. As noted in Sec.~\ref{sec:cross_section_analysis}, the specific numerical partition between $\sigma_\mathrm{B}$ and $\sigma_\mathrm{int}$ is basis-dependent, but their sum $\sigma_\mathrm{B} + \sigma_\mathrm{int}$ (the total continuum contribution) and the qualitative conclusions regarding coherent interference remain basis-independent.

\subsection{Absorption Mechanism in Weakly Bound Nuclei: $^6$Li Case}

Building on the insights obtained from the deuteron case, we next consider the heavier and more complex projectile $^6$Li, focusing on the $^6$Li+$^{59}$Co reaction at 21.5 MeV and the $^6$Li+$^{208}$Pb reaction at 33 MeV. The $^6$Li projectile is described within an $\alpha$-$d$ cluster model. In the initial analysis, continuum states with $\ell \leq 1$ are included and discretized up to an excitation energy of 12 MeV, which allows a transparent visualization of the absorption decomposition. To compare with the experimental data and assess the influence of the $d$-wave continuum, we then present a more realistic calculation for the $^6$Li+$^{208}$Pb system, extending the model space to $\ell \leq 2$ with eight bin states for each partial wave while keeping the same 12 MeV excitation-energy cutoff. For the $^6$Li+$^{59}$Co system, the fragment-target optical potentials are taken from Refs.~\cite{alpha1,yyq06}. For the $^6$Li+$^{208}$Pb system, the fragment-target optical potentials are taken from Refs.~\cite{alpha2,yyq06}; in addition, the surface part of the deuteron-target optical potential is removed following Refs.~\cite{jin15-removingsur,jin17-removingsur} for $^6$Li+$^{208}$Pb system.

\textit{Numerical robustness.} To ensure that the qualitative conclusions are not artifacts of discretization, we have verified convergence with respect to the key model-space parameters: number of bins, maximum excitation energy, maximum partial wave $\ell_{\max}$, and integration parameters (matching radius $R_{\max}$ and radial step size). Varying the number of bins by $\pm 20\%$ or extending $E_{\max}$ by 4~MeV changes $\sigma_\mathrm{A}$ by less than 3\% and leaves the ratio $\sigma_\mathrm{int}/\sigma_\mathrm{A}$ stable to within 10\%. The pattern of large negative $\sigma_\mathrm{int}$ and enhanced component cross sections in the full calculation persists across all tested discretization schemes, confirming that these are robust physical features rather than numerical artifacts.

\begin{figure}[t]
    \centering
    \includegraphics[width=1.0\linewidth]{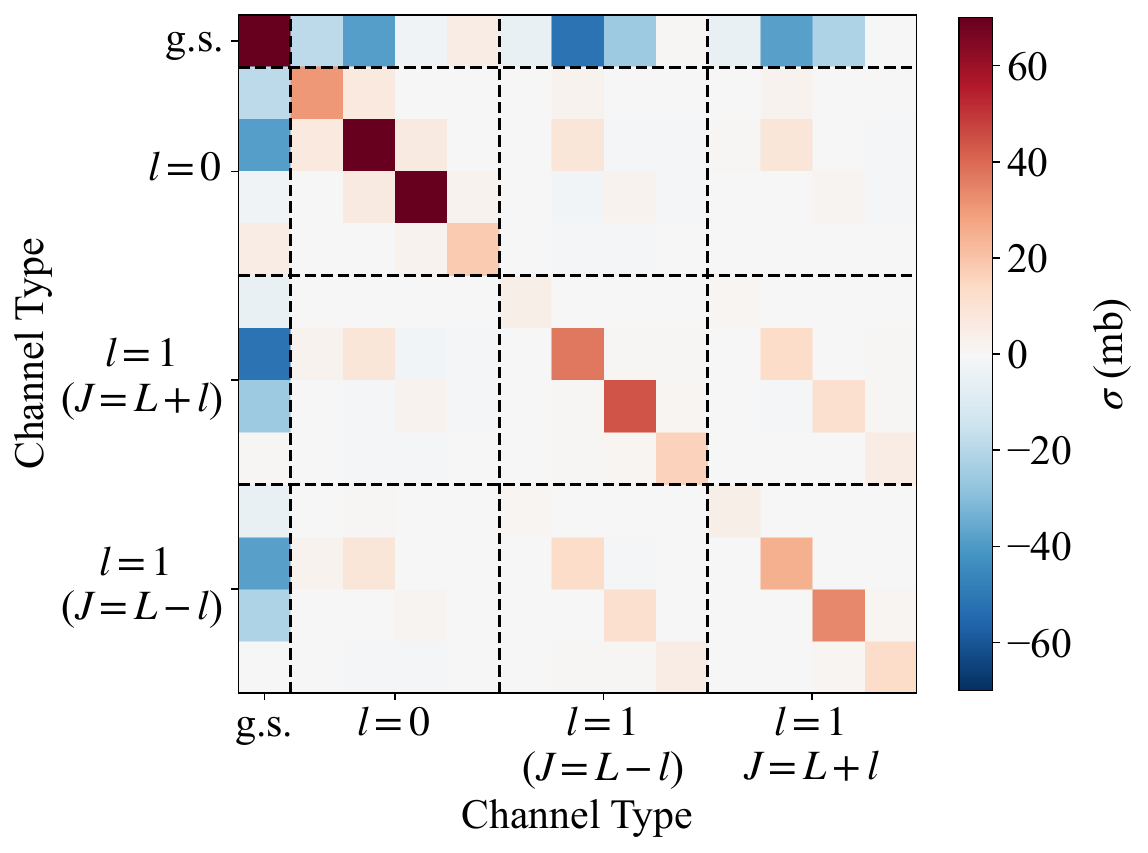}
    \caption{Decomposition of the total absorption cross section $\sigma_\mathrm{A}$ for $^6$Li+$^{59}$Co at 21.5 MeV. Red indicates positive contributions, while blue indicates negative contributions.}
    \label{fig:ab_li6}
\end{figure}

\begin{figure}[t]
    \centering
    \includegraphics[width=1.0\linewidth]{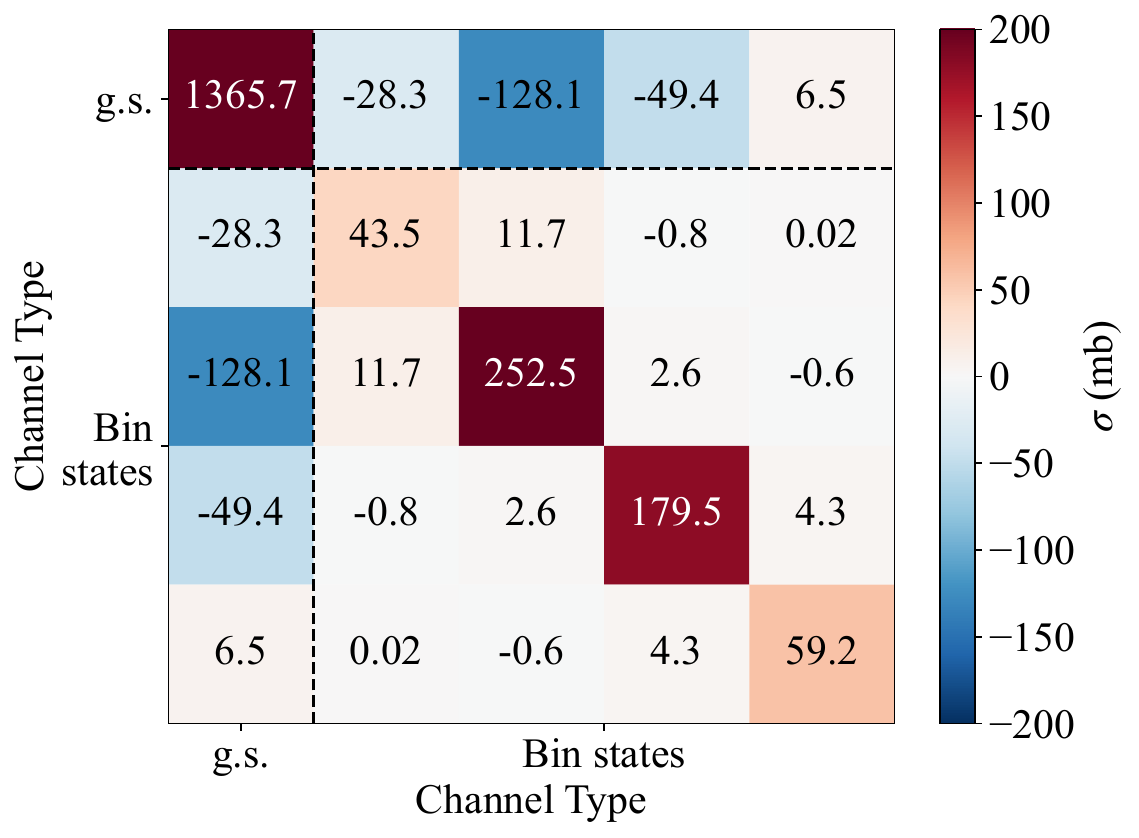}
    \caption{Decomposition in the different bin energies of the total absorption cross section $\sigma_\mathrm{A}$ for $^6$Li+$^{59}$Co at 21.5 MeV. Red indicates positive contributions, while blue indicates negative contributions.}
    \label{fig:ab_li6_bins}
\end{figure}

\begin{figure}[t]
    \centering
    \includegraphics[width=1.0\linewidth]{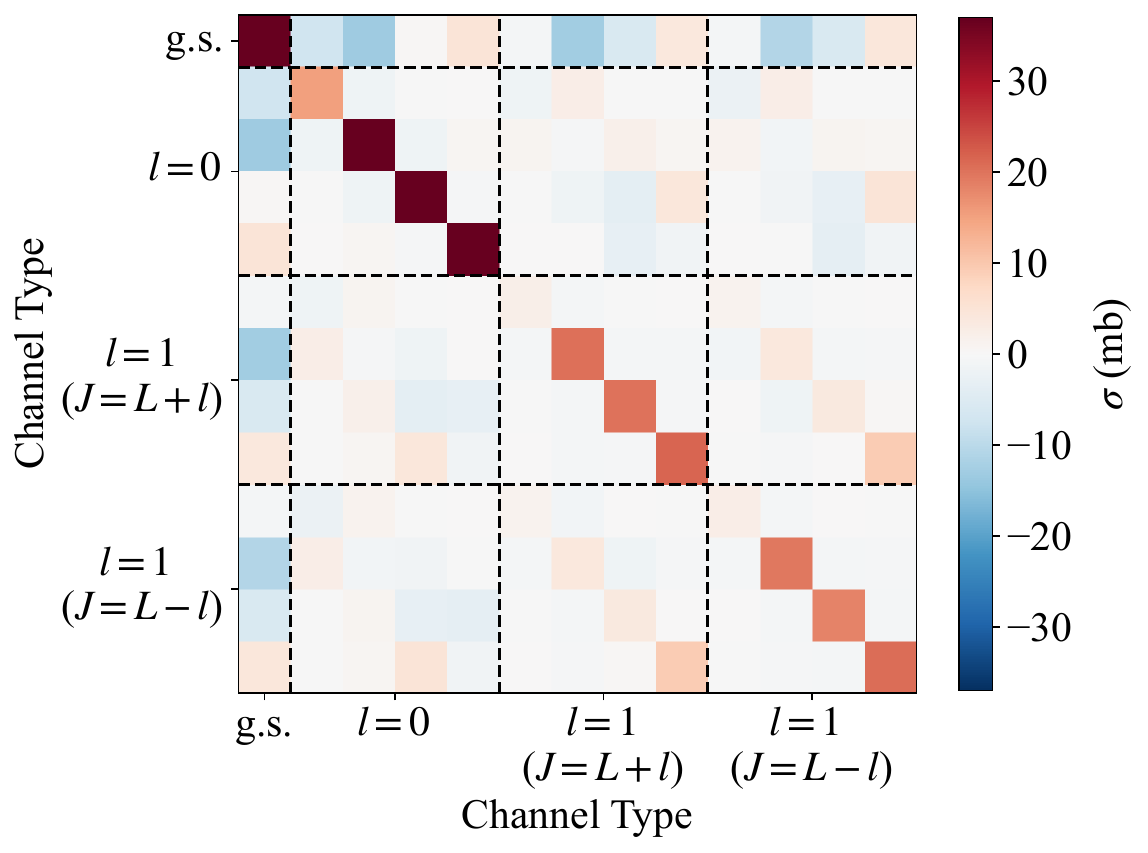}
    \caption{Decomposition of the total absorption cross section $\sigma_\mathrm{A}$ for $^6$Li+$^{208}$Pb at 33 MeV.}
    \label{fig:ab_li6Pb208}
\end{figure}

Figure~\ref{fig:ab_li6} visualizes the absorption cross-section decomposition for the $^6$Li+$^{59}$Co system. The heatmap reveals a distinct topological structure: the diagonal elements (red) representing direct absorption ($\sigma_\mathrm{D}$) and breakup absorption ($\sigma_\mathrm{B}$) are universally positive, whereas the off-diagonal elements (blue) representing the interference term ($\sigma_\mathrm{int}$) are predominantly negative. This pattern is particularly pronounced for the $l=1$ continuum states. Physically, the $l=1$ ($p$-wave) continuum in $^6$Li corresponds to the $\alpha$-$d$ relative motion with one unit of angular momentum, which couples strongly to the $l=0$ ground state through the dipole component of the fragment--target interaction. This strong coupling, combined with the lower centrifugal barrier for $l=1$ compared to higher partial waves, produces large off-diagonal matrix elements $W_{1j}$ and hence dominant interference contributions. The energy-resolved perspective in Fig.~\ref{fig:ab_li6_bins} further confirms that this interference is not an artifact of energy averaging; rather, the negative $\sigma_\mathrm{int}$ contribution persists across the excitation energy spectrum, actively competing with the positive breakup absorption terms. A qualitatively similar pattern is observed for the heavier target system, $^6$Li+$^{208}$Pb (Figure~\ref{fig:ab_li6Pb208}), indicating that this interference mechanism is a robust feature of weakly bound projectile reactions, independent of the target mass.

To quantify the dynamical role of the ground state-continuum couplings, we compare the full calculation with the $W_{1j}=0$ approximation in Table~\ref{tab:w1j_impact_li6}. The results reveal a profound flux redistribution mechanism that is consistent between the Co and Pb targets but far more dramatic in magnitude than the deuteron case. In the full calculation, the interference term $\sigma_\mathrm{int}$ provides a large negative contribution ($-398.5$~mb for Co and $-87.4$~mb for Pb with $\ell \leq 1$), which acts to regulate the total absorption cross section. Consequently, neglecting these couplings leads to an overestimation of $\sigma_\mathrm{A}$ by approximately 10.7\% for the Co system and 5.6\% for the Pb system (in the simplified $\ell \leq 1$ basis).

\begin{table}[htbp]
\centering
\caption{Impact of neglecting $W_{1j}$ on absorption cross sections for the ${}^6$Li system ($\ell \leq 1$).}
\label{tab:w1j_impact_li6}
\begin{tabular}{lccc}
\hline\hline
Quantity & Full $W_{ij}$ & $W_{1j}=0$ & Diff. \\
\hline
\multicolumn{4}{c}{${}^6$Li+${}^{59}$Co at 21.5 MeV} \\
\hline
$\sigma_\mathrm{D}$ [mb] & 1365.66 & 1338.63 & $+27.03$ \\
$\sigma_\mathrm{B}$ [mb] & 569.19 &  362.08 & $+207.11$ \\
$\sigma_\mathrm{int}$ [mb] & $-398.53$ & $0.0$ & $-398.53$ \\
$\sigma_\mathrm{A}$ [mb] & 1536.31 & 1700.71 & $-164.40$ \\
\hline
\multicolumn{4}{c}{${}^6$Li+${}^{208}$Pb at 33 MeV} \\
\hline
$\sigma_\mathrm{D}$ [mb] & 544.03 & 520.77 & $+23.26$ \\
$\sigma_\mathrm{B}$ [mb] & 324.42 &  304.15 & $+20.27$ \\
$\sigma_\mathrm{int}$ [mb] & $-87.4$ & $0.0$ & $-87.4$ \\
$\sigma_\mathrm{A}$ [mb] & 781.06 & 824.92 & $-43.86$ \\
\hline\hline
\end{tabular}
\end{table}

However, the most striking physical insight lies in the decomposition of the absorption components. The full calculation shows significantly higher values for both direct absorption ($\sigma_\mathrm{D}$) and breakup absorption ($\sigma_\mathrm{B}$) compared to the approximation. For the $^6$Li+$^{59}$Co system, the full coupling enhances $\sigma_\mathrm{B}$ by 207.1~mb (from 362.1 to 569.2~mb, a 57\% increase). For the $^6$Li+$^{208}$Pb system (with $\ell \leq 1$), $\sigma_\mathrm{B}$ increases by 20.3~mb (from 304.2 to 324.4~mb) and $\sigma_\mathrm{D}$ by 23.3~mb. This implies that the off-diagonal imaginary couplings act as a ``pump,'' facilitating the transfer of flux from the elastic channel into the breakup absorption channels. When $W_{1j}$ is removed, this bridge is broken, starving the breakup absorption channels of flux.

\begin{figure}[t]
    \centering
    \includegraphics[width=1.0\linewidth]{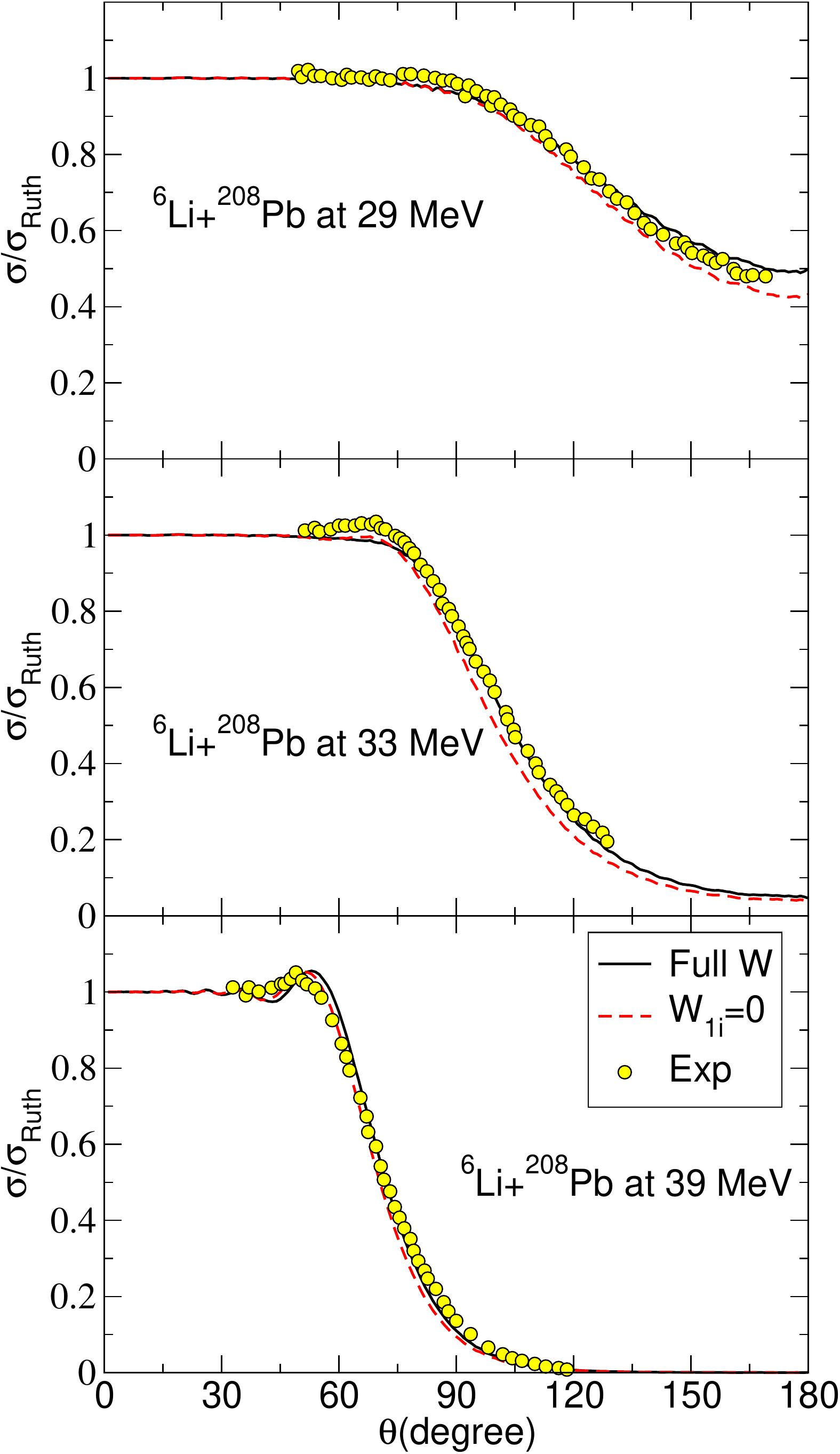}
    \caption{Elastic scattering angular distributions for $^6\mathrm{Li}+{}^{208}\mathrm{Pb}$ at 29 MeV, 33 MeV, and 39 MeV, shown as the ratio $\sigma/\sigma_\mathrm{Ruth}$. CDCC calculations with $\ell \leq 2$ are compared: full coupling (solid line) and the $W_{1j}=0$ approximation (dashed line), experimental data (yellow circles)~\cite{KEELEY1994326}.}
    \label{fig:elastic_li6}
\end{figure}

To assess the role of the $d$-wave ($\ell=2$) continuum states and to examine the energy dependence of the coupling effects, we present in Table~\ref{tab:w1j_impact_li6Pb208} calculations for $^6$Li+$^{208}$Pb employing $\ell \leq 2$ with 8 bin states per partial wave at three incident energies: 29, 33, and 39 MeV. The corresponding elastic scattering angular distributions are shown in Fig.~\ref{fig:elastic_li6}.

At 33 MeV, the extension to $\ell = 2$ substantially amplifies the flux redistribution compared to the $\ell \leq 1$ basis: $\sigma_\mathrm{B}$ increases from 324.4~mb to 453.2~mb (a 40\% enhancement) and the magnitude of the interference term deepens from $-87.4$ to $-222.1$~mb, while the total absorption $\sigma_\mathrm{A}$ remains nearly unchanged at $\approx 779$~mb, confirming flux conservation. The overestimation of $\sigma_\mathrm{A}$ when setting $W_{1j}=0$ thereby grows from 5.6\% (with $\ell \leq 1$) to 11.5\% (with $\ell \leq 2$), underscoring that any incoherent approximation becomes progressively less reliable as the model space is enlarged to include the physically important $\ell=2$ continuum.

The multi-energy comparison in Table~\ref{tab:w1j_impact_li6Pb208} reveals a systematic energy dependence of the coupling effects. As the incident energy increases from 29 to 39 MeV, the absolute magnitude of $\sigma_\mathrm{int}$ grows substantially (from $-64.8$ to $-222.1$ to $-405.2$~mb), and the full-coupling calculation enhances both $\sigma_\mathrm{B}$ (by 8\%, 20\%, and 23\% at 29, 33, and 39~MeV, respectively) and $\sigma_\mathrm{D}$ (by 1.5\%, 11.6\%, and 18.3\%) relative to the $W_{1j}=0$ approximation. This indicates that the off-diagonal imaginary couplings play an increasingly important dynamical role in redistributing flux into specific absorption channels as the reaction moves above the Coulomb barrier. Counterintuitively, the relative overestimation of the total $\sigma_\mathrm{A}$ by the approximation actually \emph{decreases} with energy, from 17.7\% at 29~MeV to 11.5\% at 33~MeV and 7.0\% at 39~MeV. This occurs because the overall growth of $\sigma_\mathrm{A}$ with energy outpaces the growth of the absolute overestimation, even as the internal flux redistribution becomes more pronounced.

The elastic-scattering angular distributions in Fig.~\ref{fig:elastic_li6} provide an additional perspective on the same coupling effects. The full $\ell \leq 2$ calculation reproduces the elastic scattering angular distributions~\cite{KEELEY1994326} well at all three energies, whereas the $W_{1j}=0$ approximation shows clear deviations, especially at 29 MeV. As the incident energy increases to 33 and 39 MeV, the difference between the two calculations in the elastic channel becomes progressively smaller. This indicates that near-barrier elastic scattering is particularly sensitive to the detailed structure of the imaginary coupling matrix, while at higher energies the dominance of strong overall absorption reduces that sensitivity.
\begin{table}[htbp]
\centering
\caption{Impact of neglecting $W_{1j}$ on absorption cross sections for the ${}^6$Li+${}^{208}$Pb system with $\ell \leq 2$ (8 bin states per partial wave).}
\label{tab:w1j_impact_li6Pb208}
\begin{tabular}{lccc}
\hline\hline
Quantity & Full $W_{ij}$ & $W_{1j}=0$ & Diff. \\
\hline
\multicolumn{4}{c}{${}^6$Li+${}^{208}$Pb at 29 MeV, $\ell \leq 2$} \\
\hline
$\sigma_\mathrm{D}$ [mb] & 216.21 & 212.96 & $+5.25$ \\
$\sigma_\mathrm{B}$ [mb] & 137.56 &  127.38 & $+10.18$ \\
$\sigma_\mathrm{int}$ [mb] & $-64.75$ & $0.0$ & $-64.75$ \\
$\sigma_\mathrm{A}$ [mb] & 289.03 & 340.34 & $-51.31$ \\
\hline
\multicolumn{4}{c}{${}^6$Li+${}^{208}$Pb at 33 MeV, $\ell \leq 2$} \\
\hline
$\sigma_\mathrm{D}$ [mb] & 548.23 & 491.33 & $+56.90$ \\
$\sigma_\mathrm{B}$ [mb] & 453.20 &  377.45 & $+75.75$ \\
$\sigma_\mathrm{int}$ [mb] & $-222.1$ & $0.0$ & $-222.10$ \\
$\sigma_\mathrm{A}$ [mb] & 779.33 & 868.78 & $-89.45$ \\
\hline
\multicolumn{4}{c}{${}^6$Li+${}^{208}$Pb at 39 MeV, $\ell \leq 2$} \\
\hline
$\sigma_\mathrm{D}$ [mb] & 984.43 & 832.04 & $+152.39$ \\
$\sigma_\mathrm{B}$ [mb] & 824.05 &  670.04 & $+154.01$ \\
$\sigma_\mathrm{int}$ [mb] & $-405.16$ & $0.0$ & $-405.16$ \\
$\sigma_\mathrm{A}$ [mb] & 1403.32 & 1502.08 & $-98.76$ \\

\hline\hline
\end{tabular}
\end{table}

These findings challenge the validity of incoherent approximations for weakly bound nuclei. The interference couplings play a dual role: they enhance specific reaction pathways (boosting $\sigma_\mathrm{D}$ and $\sigma_\mathrm{B}$), while simultaneously contributing a negative interference term that reduces the total reaction probability. Thus, retaining the full non-diagonal imaginary coupling matrix is necessary for correctly predicting the partitioning of flux between direct and breakup mechanisms.

\section{Summary and Discussion}

In this work, we have applied the generalized optical theorem~\cite{COTANCH201048} within the Continuum-Discretized Coupled-Channels (CDCC) framework and investigated the role of off-diagonal imaginary couplings in the absorption mechanism of weakly bound nuclei. By decomposing the total absorption cross section into direct absorption ($\sigma_\mathrm{D}$), breakup absorption ($\sigma_\mathrm{B}$), and the interference term ($\sigma_\mathrm{int}$), we have elucidated the complex quantum interference effects that govern the redistribution of reaction flux.

The theoretical analysis establishes that the total absorption is not merely an incoherent sum of channel probabilities. Instead, it includes coherent interference terms ($\sigma_\mathrm{int}$) arising from the off-diagonal imaginary couplings ($W_{ij}$). Our numerical calculations for d+$^{93}$Nb and $^6$Li+$^{59}$Co/$^{208}$Pb consistently demonstrate that these interference terms are predominantly negative in the systems studied, representing a destructive interference mechanism that plays a central role in flux conservation within the model space.

The comparison between full coupled-channel calculations and the $W_{1j}=0$ approximation reveals a dual role of the off-diagonal imaginary couplings: they enhance specific absorption channels while generating a negative interference term. In all systems studied, the full calculation increases both $\sigma_\mathrm{B}$ and $\sigma_\mathrm{D}$ relative to the $W_{1j}=0$ limit, with the effect becoming particularly strong for $^6$Li+$^{59}$Co and for $^6$Li+$^{208}$Pb once the $\ell=2$ continuum is included. At 33 MeV for $^6$Li+$^{208}$Pb, extending the model space from $\ell \leq 1$ to $\ell \leq 2$ enhances $\sigma_\mathrm{B}$ by 40\% and deepens $\sigma_\mathrm{int}$ from $-87.4$ to $-222.1$~mb, showing that the coupling effects grow with the size of the model space.

In all systems studied, $\sigma_\mathrm{int}$ is large and negative, acting as a regulator in the coherent decomposition. Neglecting it therefore overestimates the total absorption cross section, by about 10\% for $^6$Li+$^{59}$Co, by 5.6\% for $^6$Li+$^{208}$Pb at 33 MeV with $\ell \leq 1$, and by 11.5\% for the corresponding $\ell \leq 2$ calculation. For $^6$Li+$^{208}$Pb with $\ell \leq 2$, the magnitude of $\sigma_\mathrm{int}$ increases strongly from 29 to 39 MeV, while the relative overestimation of $\sigma_\mathrm{A}$ decreases from 17.7\% to 7.0\%. Thus, flux redistribution becomes stronger with increasing energy even though the relative error in $\sigma_\mathrm{A}$ becomes smaller. We note that the sign of $\sigma_\mathrm{int}$ is an empirical result of the present calculations; whether destructive interference is a general property of weakly bound systems remains an open theoretical question.

This analysis clarifies why models that neglect off-diagonal imaginary couplings are physically incomplete. We have also verified that the qualitative conclusions are numerically robust against reasonable changes in binning, excitation cutoff, maximum continuum partial wave, and radial integration parameters.

A potential objection is that the comparison between full and $W_{1j}=0$ calculations in this work employs identical optical potential parameters. One might argue that re-fitting the parameters within the $W_{1j}=0$ approximation could restore agreement with elastic scattering data, effectively ``renormalizing'' the missing $\sigma_\mathrm{int}$ physics into adjusted potential depths or geometries. While such parameter adjustment may indeed recover acceptable fits to elastic angular distributions, it comes at the cost of distorting the extracted component cross sections. The re-fitted parameters would compensate for the missing coherent interference by artificially modifying the absorption strengths, leading to systematically biased values of $\sigma_\mathrm{D}$ and $\sigma_\mathrm{B}$. This is precisely the core message of the present work: merely reproducing experimental observables is insufficient; the underlying reaction mechanism must be correctly captured. A model that fits data through parameter tuning but misrepresents the flux partitioning will yield misleading physical interpretations, particularly when extrapolating to unmeasured observables or different reaction systems.

Direct experimental validation of individual components, and specifically the interference term $\sigma_\mathrm{int}$, is intrinsically impossible: detectors register asymptotic products, not internal coherence terms of the wave function. Component extractions (\textit{e.g.}, direct vs. breakup fractions) are therefore model-dependent. Moreover, we emphasize that the absorption cross section $\sigma_\mathrm{A}$ calculated within the CDCC framework includes all flux absorbed by the imaginary potentials, encompassing not only fusion but also transfer reactions, deep inelastic scattering, and other non-elastic channels. The quantities $\sigma_\mathrm{D}$ and $\sigma_\mathrm{B}$ should be interpreted as components of total absorption rather than pure fusion. Disentangling genuine fusion from transfer would require either fusion-specific short-range potentials or explicit coupling to transfer channels. Nevertheless, the central conclusion, that coherent interference generates a significant $\sigma_\mathrm{int}$ term (observed to be negative in the systems studied), applies to any absorption mechanism governed by non-diagonal imaginary couplings. The value of the present work is to provide a rigorous theoretical constraint (the generalized optical theorem and its component analysis) based on flux conservation.

\textit{Implications for experimental analysis.} Our findings have direct consequences for the extraction of reaction cross sections from experimental data. A common experimental procedure is to measure the total fusion cross section and then use theoretical models to partition it into complete fusion (direct) and incomplete fusion (breakup-induced) components. If the theoretical model employed for this partition neglects the off-diagonal imaginary couplings, as is done in many simplified CDCC or optical model analyses, the extracted breakup contribution will be systematically underestimated. We therefore urge experimentalists to adopt full-coupling CDCC calculations, retaining all $W_{ij}$ terms, as the baseline for any model-assisted decomposition of absorption data (which may be identified with fusion when the imaginary potential is designed specifically for fusion absorption). Furthermore, when comparing theoretical predictions with measured elastic scattering angular distributions, the full-coupling calculation should be used; as shown in Figs.~\ref{fig:elastic} and \ref{fig:elastic_li6}, simplified models produce qualitatively different angular distributions and can be especially misleading near the Coulomb barrier, where the elastic channel is most sensitive to the detailed structure of the imaginary coupling matrix.


In conclusion, the component analysis presented here demonstrates that accurate predictions for weakly bound nuclei require the retention of the full non-diagonal imaginary coupling matrix. The quantum interference mediated by these couplings is not merely a perturbative correction but a significant driver of reaction dynamics in the systems studied. It determines the correct partitioning between direct and breakup mechanisms while respecting flux conservation, and its impact depends both on the adopted model space and on the incident energy. Beyond CDCC, the generalized optical-theorem-based component analysis is applicable to broader coupled-channel problems (including transfer), and we advocate it as a standard paradigm for mechanism-resolved extraction of reaction observables.

\begin{acknowledgments}
We thank Stephen R. Cotanch for developing the generalized coupled-channels optical theorem employed in this work and for insightful comments that improved the presentation of the manuscript. We also thank Antonio M. Moro for useful discussions.
This work was supported by the National Natural Science Foundation of China (Grant Nos. 12535009 and 12475132), the National Key R\&D Program of China (Contract No. 2023YFA1606503), and the Fundamental Research Funds for the Central Universities.
\end{acknowledgments}

\bibliography{inclusive}

\end{document}